\documentclass[12pt]{article}
\usepackage{graphicx,psfrag,epsfig}
\usepackage{amssymb,amsmath,amscd,amsthm,mathtools}
\usepackage{graphicx,psfrag,epsfig}
\usepackage[english]{babel}
\usepackage{listings}

\usepackage{float}

\usepackage{graphicx}
\usepackage{pstricks}
\usepackage{psfrag}
\usepackage{csquotes}
\usepackage{cite}

\usepackage{comment}

\usepackage{hyperref}

\usepackage[left=2cm,right=2cm,top=2cm,bottom=2cm,bindingoffset=0cm]{geometry}

\usepackage{hyperref}
\hypersetup{
  colorlinks   = true,          %Colors links instead of ugly boxes
  urlcolor     = violet,          %Color for external hyperlinks
  linkcolor    = blue,          %Color of internal links
  citecolor   = violet             %Color of citations
}
\hypersetup{linktocpage}

\newtheorem{alg}{Algorithm}

\DeclareMathOperator{\rel}{rel}

\mathtoolsset{showonlyrefs}

\title{Modifications to a classic BFGS library for use with SIMD-equipped hardware and an AAD library}
\author{Evgeny Goncharov\thanks{Department of Pure Mathematics and Mathematical Statistics, University of Cambridge, UK., and MatLogica, London, UK., eg555@cam.ac.uk,   evgeny.goncharov@matlogica.com}, Alexandre Rodrigues\thanks{Department of Physics, University of Aveiro, Portugal, alexandrerrodrigues@ua.pt}}
\date{}

\begin{document}

\maketitle

\begin{abstract}
We introduce certain modifications of the BFGS method for functions that are not parallelizable by nature (having consecutive operations only) taking advantage of SIMD. We also provide a modified LBFGS\texttt{++} library that takes advantage of these modifications, and the use of AAD, and give an interface for AAD users that takes advantage of the modified library automatically. We give two examples to illustrate the performance. The modified library is up to 3.8 times faster for European Swaption curve calibration in ORE\cite{ore} (not parallelizable) and 1.4 times faster for calibrating the LMM model by a set of European options. 
\end{abstract}

\tableofcontents

\section{Introduction}

	The classical Newton’s method \cite{num_analysis} uses the Hessian matrix $H$ (the matrix of second derivatives) to find an extremum of a function $f: \mathbb{R}^n \to \mathbb{R}$ by an iterative algorithm starting at some initial point $x_0$. However, computing the Hessian can be too expensive, which gave rise to the development of the less costly \textit{quasi-Newton} methods. The first one was proposed by William C. Davidon in \cite{davidon} but it is rarely used now as more efficient methods have been developed. Popular methods today include the symmetric rank-one formula \cite{sr1} and the BHHH method \cite{bhhh}. The BFGS method \cite{broyden,fletcher,golfarb,shanno} and its low-memory extension LBFGS \cite{lbfgs} are currently the most used. We give an overview of the algorithms in Section \ref{bfgsalg}. These methods require little memory, have low computational complexity since no matrix inversions are required, and can find the solution even when the initial point $x_0$ is quite far from the extremum. 
	
	Parallelization is an important tool to improve computing efficiency. The two main approaches to parallelization are \textit{multi-threading} and \textit{Single Instruction Multiple Data (SIMD)}.
	Multi-threading separates the calculation into parallel independent computations (threads) whereas SIMD allows taking multiple data points as input and processing them at the same time. It is straightforward to implement multi-threading into the BFGS method for a parallelizable function $f$. However, if $f$ is not parallelizable (e.g. it is a sequence of consecutive operations), it is less clear how to use parallelization to enhance performance. We review the BFGS method and introduce our parallelization modifications utilizing SIMD in Section \ref{modify}. We use Advanced Vector Extensions (AVX) \cite{avx} in the process. 
	
	We implemented the modifications into the LBFGS\texttt{++} library, the modified library is available at \cite{finalgit}. The interface of the classical library assumes a functor (an AAD library) that calculates both the value (forward mode) and the gradient (reverse mode)  of the target function $f$. However, straightforward use of an AAD library may lead to unnecessary computations. We modify the functions in the library to avoid them, see Section \ref{derivatives}.
	
	The AAD library that we use is Matlogica's AADC \cite{aadc_manual}. It is compatible with AVX and SIMD which makes it ideal for the project. We developed a C\texttt{++} class that allows AADC users to utilize the (modified version of) LBFGS\texttt{++} library automatically. 
	
	We discuss implementation in Section \ref{impl} and give two examples to illustrate performance. In the first example, we take the Open Source Risk Project (ORE) \cite{ore} code for the European Swaption curve calibration and pass it through the AADC library to get optimized and vectorized binary kernels for the forward (computing the function) and the reverse (computing the gradient) passes of $f$. We achieve an up to 3.8 times acceleration for the optimized code with the modified library. In the second example, we calibrate the Libor Market Model (LMM) \cite{lmm_art} using the recipe of \cite{gl} and achieve a 1.4 times acceleration with the modified library. We note that in this case all the acceleration is achieved just by avoiding the unnecessary gradient computations mentioned above. Note also that the recent \cite{bgg} gives alternative recipes for performing AAD in this setup. All the code is available at \cite{finalgit}.
	
	Finally, we discuss other possible approaches to improving the performance of the LBFGS\texttt{++} library in Section \ref{concl}.

\textbf{Acknowlegements.} We are grateful to Evgeny Lakshtanov for the useful advice, discussions, and the suggestion to write up these ideas. We are also thankful to Dmitri Goloubentsev for the idea to use polynomial fitting (Section \ref{polyfit}), and to Matlogica LTD for providing access to the AADC library.

\section{The BFGS method} \label{bfgsalg}

The BFGS method \cite{broyden,fletcher,golfarb,shanno} for computing a minimum of a function $f: \mathbb{R}^n \to \mathbb{R}$ runs as follows.
    
    %% Give a brief description of the algorithm like on Wiki but so that it is clear what the $x_k, \alpha_k, p_k$ are.
    
    % Do some number of line searches. Stop each line search when certain termination conditions are reached, list them, cite for more info. Explain clearly how we update between line searches. Does $x_k$ change here between linesearches? $p_k$ stays the same direction during line search?
    
    \begin{alg}[BFGS] \label{bfgs}
    \begin{enumerate}
    Start with an initial guess $x_0$ and a Hessian matrix estimation $B_0$. Repeat the following steps for $k \geq 0$.
        \item Obtain the search direction: $p_k := -B_k^{-1} \nabla f(x_k)$.
        \item Perform a line search (one-dimensional optimization, see \cite{nw} for details) to find the acceptable step $\alpha_k$ in the direction $p_k$. Line search is terminated when a chosen termination condition (Armijo, Wolfe, or Strong Wolfe) is reached. 

        Line search produces a sequence of steps $\alpha_k^i$ for $i \geq 0$ and $x_k^i: = x_k^{i-1} + \alpha_k^i p_k$ for $k \geq 1$ (here $x_k^0:=x_k$) until an acceptable step $\alpha_k$ is found (the termination condition is reached). 
        \item Use the displacement $s_k := \alpha_k p_k $ to update the variable $x_{k+1} := x_k + s_k$.
        \item Set the change of gradient $y_k := \nabla f(x_{k+1}) - \nabla f(x_k) $.
        \item Update the Hessian matrix estimation $$ B_{k+1}:=B_{k}+{\frac{ {y}_{k} {y}_{k}^{\mathrm{T}} }{{y}_{k}^{\mathrm{T}} {s}_{k}}} - {\frac{B_{k} {s}_{k} {s}_{k}^{\mathrm {T}} B_{k}^{\mathrm {T} }}{{s}_{k}^{{T}}B_{k} {s}_{k}}}.$$
    \end{enumerate}
    
    As $k \to \infty$, the $x_k$ converge to a minimum of $f$. 
    \end{alg}
    
    % Explain what LBFGS means, mention two modes of the LBFGS\texttt{++} library, mention the additional parameters that we use later.
    The LBFGS algorithm is an enhancement of Algorithm \ref{bfgs} that has similar steps with certain modifications that allow to store less data in the memory (see \cite{lbfgs} for details). LBFGS\texttt{++} is a classical library implementing the BFGS and LBFGS methods. 
    We make use of certain LBFGS\texttt{++} parameters given to the solver in one \verb*|LBFGSParam| object:
	\begin{itemize}
		\item $\varepsilon_{\rel}$: we change it to guarantee convergence in some applications;
		\item \verb*|max_iterations|: limited to 100 to reduce computational time in LMM (Section \ref{lmm});
		\item \verb*|linesearch|: we tested for various line search termination conditions.
	\end{itemize}

\section{Modifications to the BFGS method and the LBFGS\texttt{++} library} \label{modify}

    We shall now explain our modifications to the BFGS algorithm, and the LBFGS\texttt{++} library. All our modifications only concern line search, that is Step 2 of Algorithm \ref{bfgs}.

\subsection{Parallelizing line search (LS)} \label{parlinesearch}
    \label{subsec:parallelLinesearch}

    Suppose that we are at the line search step of Algorithm \ref{bfgs} and we are about to compute $f(x_k^i) =f(x_{k}^{i-1} + \alpha_k^i p_k)$. A simple way to improve performance is to instead calculate the function value at several points at each call of the functor and choose the point with the smallest value of the function.
    
 	We use AVX as it allows us to use $4$ points (AVX-2) or $8$ points (AVX-512) and is well-integrated into AADC. For AVX-2 we use the $4$ points:
 	
 	\begin{equation*}
 		x_{k}^{i-1} + \frac{1}{2}\alpha_k^i p_k,\ x_{k}^{i-1} + \alpha_k^i p_k,\ x_{k}^{i-1} + \frac{3}{2}\alpha_k^i p_k, \ x_{k}^{i-1} + 2\alpha_k^i p_k.
 	\end{equation*}
 	
 	For AVX-512 we use the 8 points:
 	\begin{equation*}
 		\begin{split}
 			&x_{k}^{i-1} + \frac{1}{4}\alpha_k^i p_k,\ x_{k}^{i-1} + \frac{1}{2}\alpha_k^i p_k,\ x_{k}^{i-1} + \frac{3}{4}\alpha_k^i p_k, \ x_{k}^{i-1} + \alpha_k^i p_k, \\
 			&x_{k}^{i-1} + \frac{5}{4}\alpha_k^i p_k,\ x_{k} + \frac{3}{2}\alpha_k^i p_k,\ x_{k}^{i-1} + \frac{7}{4}\alpha_k^i p_k, \ x_{k}^{i-1} + 2\alpha_k^i p_k.
 		\end{split}	
 	\end{equation*}
 	
 	We let $x_{k}^i$ be the point for which the value of $f$ is the smallest among the $4$ or $8$ points respectively, and proceed with line search (or do a Hessian update and proceed with the algorithm if the termination condition has been reached).
 	This modification alone results in acceleration of the European Swapton curve calibration by a factor of $2$ as we see in Section \ref{ore}.  

\subsection{Polynomial regression (Polyfit)} \label{polyfit}

\begin{comment}
https://gist.github.com/ksjgh/4d5050d0e9afc5fdb5908734335138d0
It is a least-square polynomial fitting 
https://towardsdatascience.com/least-square-polynomial-fitting-using-c-eigen-package-c0673728bd01
Maybe I should make it more readable

        VectorXd polyfit(VectorXd xvals, VectorXd yvals, int order) 
    {
        Eigen::MatrixXd A(xvals.size(), order + 1);

        for (int i = 0; i < xvals.size(); i++)
            A(i, 0) = 1.0;

        for (int j = 0; j < xvals.size(); j++) {
            for (int i = 0; i < order; i++)
                A(j, i + 1) = A(j, i) * xvals(j);
        }

        auto Q = A.householderQr();
        auto result = Q.solve(yvals);
        return result;
    }
\end{comment}

% \textcolor{green}{explain (?)} \textcolor{red}{I'd rather give some explanation why we use this or if some popular interpolation method is used - name the method}

        We proceed with the same setup as above. However, instead of letting $x_k^i$ be the point where $f$ is the smallest, we use polynomial fitting as follows. We fit the function values at the $4$ or $8$ points using a polynomial of degree $3$ or $7$ respectively as a least-square polynomial fitting using the QR factorization from the Eigen library \cite{eigen}. This gives a polynomial:
	 	\begin{align*}
	 		f(\alpha) &= p_0 + p_1 \alpha + p_2 \alpha^2 + p_3 \alpha^3 \text{ (if using AVX-2)}\\
	 		f(\alpha) &= p_0 + p_1 \alpha + p_2 \alpha^2 + p_3 \alpha^3 + p_4 \alpha^4 + p_5 \alpha^5 + p_6 \alpha^6 + p_7 \alpha^7 \text{ (if using AVX-512)}
	 	\end{align*}
	 	and we can find its roots. 
	 	Let the smallest root be $\alpha_k^{\min}$ and set $x_{k}^{\min} :=  x_k^{i-1} + \alpha_{\min} p_k$.

        We found the best acceptable step interval to be $\alpha_k^{\min} \in [\frac{1}{4}\alpha_k , 4\alpha_k]$. 
        If $\alpha_{\min} \in [\frac{1}{4}\alpha_k , 4\alpha_k]$ and $f(x_k^{\min}) < f(x_k^{i-1})$ we let $x_k^i:=x_k^{\min}$. Otherwise, we use the $x_k^i$ of Section \ref{parlinesearch}.
	 	This turns out to be a more powerful modification than the one of  Section \ref{parlinesearch} and offers an acceleration of the European Swapton curve calibration by a factor of $3.8$ (again, see Section \ref{ore}).  

\subsection{Computing derivatives} \label{derivatives}

    By studying the line search implementation of the LBFGS\texttt{++} library we realized that we can improve performance by separating the interface of the \verb*|operator()| function into 3 different functions. The \verb*|operator()| function of LBFGS\texttt{++} calculates both the target function value and the gradient. However, none of the termination conditions use the gradient which is only required by Step 4 of the LBFGS version of Algorithm \ref{bfgs}. The termination conditions only require directional derivatives at Step 2 of the LBFGS version of Algorithm \ref{bfgs}.
    		
    Using \verb*|operator()|, the directional derivative is computed via the dot product $$\frac{\partial f}{\partial \alpha_k^i} = \nabla f(x_k^i) \cdot p_k.$$ We instead use the finite difference method implemented in a new function \verb*|getDg()|. This change allows us to move the gradient computation from \verb*|operator()| to a new function \verb*|getGrad()|, which is only called after line search.
    
    We now explain how \verb*|getDg()| works. Suppose that we are at some stage $x_k^i$ of line search. To calculate the directional derivative of $f$ at $x:=x_k^i$ we can use

% 		std::vector<double> coeffs_256 = {1/12.0,-2/3.0,2/3.0,-1/12.0};
%     std::vector<double> coeffs_512 = {1/280.0,-4/105.0,1/5.0,-4/5.0,4/5.0,-1/5.0,4/105.0,-1/280.0};
    
	% 1 ponto
	
	\begin{equation*}
		\label{eq:dg_2}
		f'(x) \approx \frac{f(x + hp_k) - f(x)}{h},
	\end{equation*}
	where $h > 0$ is a parameter that one can set using an interface parameter \verb*|delta|. This approach is simple but can cause a significant error for some target functions $f$ and parameters $h$.	
	
	% 4/8 pontos:
	A better approach is to use AVX as before and define 4 or 8 points as follows:
	
	\begin{equation*}
		x_{-2} = x-2h p_k,\ x_{-1}= x-h p_k,\ x_1 = x+h p_k,\ x_2 = x+2h p_k \text{ (if using AVX-2)}
	\end{equation*}
	\begin{multline*}
		\ \ \ \ \ \ \ \ \ x_{-4} = x-4h p_k,\ x_{-3} = x-3h p_k,\ x_{-2} = x-2h p_k, \ x_{-1} = x-h p_k,\ x_1 = x+h p_k,\\ x_2 = x+2h p_k,\ x_3 = x+3h p_k,\ x_4 = x+4h p_k \text{ (if using AVX-512)}
	\end{multline*}
	
	Then one can approximate the derivative using:
	
	\begin{equation}
		\label{eq:dg_AVX}
		f'(x) \approx \frac{\sum_{i} f(x_i) c_i}{h}
	\end{equation}
	
	where $c_i$ are the central finite difference coefficients, see \cite{fdm}.
	
	In the cases of $4$ and $8$ points we get:

	\begin{equation*}
		f'(x) \approx \frac{1}{h} \left[ \frac{1}{12}f(x-2hp_k) - \frac{2}{3}f(x-hp_k) + \frac{2}{3}f(x+hp_k) - \frac{1}{12}f(x+2hp_k) \right]
	\end{equation*}
	\begin{multline*}
			\ \ \ \ \ \ \ \ \ \ \ \ f'(x) \approx \frac{1}{h} \left[ \frac{1}{280}f(x-4hp_k) - \frac{4}{105}f(x-3hp_k) + \frac{1}{5}f(x-2hp_k) - \frac{4}{5}f(x-hp_k) \right. \\
			 \left. +\frac{4}{5}f(x+hp_k) -  \frac{1}{5}f(x+2hp_k) + \frac{4}{105}f(x+3hp_k) -\frac{1}{280}f(x+4hp_k) \right]
	\end{multline*}
	
	This broader averaging usually gives better results.

\section{Implementation and testing} \label{impl}

We shall explain the changes that we made to the LBFGS\texttt{++} library, introduce an interface for AADC that takes advantage of the modified library automatically and give two examples illustrating the efficiency of our modifications. All the code is available at \cite{finalgit}. 
% \textcolor{red}{new Github project for this compatible with the paper - without Rosenbrock and with a clear readme explaining which files are relevant to 1) modified library 2) C++ class allowing AADC users to use the library 3) ORE example 4) LMM model example}

\subsection{Implementation}
    We created a C\texttt{++} class \cite{finalgit} that facilitates the use of the LBFGS\texttt{++} library by AADC users.
    The interface includes all the functions required by our implementation.
    For most implementations we only need to pass the following AADC data to the basic class: recorded function, workspace, input and output variables.
	
	One can set the value of $h$ in the computation of the directional derivative.
	The implementations that need extra data or an iterative method to calculate the gradient or function value require straightforward changes to the interface.
	By default we calculate the function value as the sum of all the outputs since most target functions are of the form $f: \mathbb{R}^n \rightarrow \mathbb{R} $.
	
	%We declare AADC data as normal and create the AADC function. Then we can then construct an object of our class and use LBFGS\texttt{++} as normal.

% 	As this work is based on AVX we can disable the line search parallelization by defining the macro \texttt{BFGS\_HELPER\_NO\_AVX} (note that this disables almost everything and should be slower than the original). The default interface uses AVX2 (4 points of computation), to use 8 we can set the macro \texttt{BFGS\_HELPER\_AVX512}.
	
	As our implementation is based on AVX we cannot disable the line search parallelization but one can set the number of computation points with the \texttt{typedef} template, i.e. initializing the interface object as \texttt{BFGSHelper<mmType>}.
	
    Polynomial regression is achieved with a polynomial of order $3$ (for AVX-2) or $7$ (for AVX-512) by default as it uses \texttt{avx\_count} points to compute it. The order can be set and verified by the functions \texttt{setPolyFitOrder(int ord)}  and \texttt{int getPolyFitOrder()}. It can also be fully disabled by setting the order to 0.

	The computation of the directional derivative uses all the available \texttt{avx\_count} points. The options are using $2, \ 4, \ 6$ or 8 points (note that for AVX2 the maximum is $4$). The user can use the functions \texttt{setDgOrder(int ord)}  and \texttt{int getDgOrder()} to set and verify the number of points used. If an invalid number is set, the procedure is disabled by substituting it with the original implementation where the directional derivative is computed as the dot product of the gradient and the direction.
	
	The user can also set the finite differences coefficients $c_i$ using the functions \\ \texttt{setDgCoeffs(std::vector<double> cs)}  and \texttt{std::vector<double> setDgCoeffs()} (although the interface already has the default central finite difference coefficients for any valid number of points).

\subsection{Curve calibration in ORE} \label{ore}

    %% Briefly explain the setup - just what example in ore we are using,
    We used the third example in the Open Source Risk Project (ORE) user guide \cite{ore} that shows the exposure evolution of European Swaptions with cash and physical delivery. This example already had AADC integrated into the underlying quantitative finance library QuantLib \cite{ql}.
    
    %% What the loss function that we want to minimize is
    We made some minor changes to \verb*|reportwriter.cpp| and used as input the curve \enquote{xois 0 EUR}. This curve has 14 outputs and 33 curve parameters so the curve function is $g: \mathbb{R}^{33} \rightarrow \mathbb{R}^{14} $. Our target function $f: \mathbb{R}^{33} \to \mathbb{R}$ for ORE is the quadratic loss function relative to the desired outputs:
	
	\begin{equation}
		f(x) := \sum_{i=1}^{14} \left( g(x)_i - g(x_{\min})_i \right)^2
	\end{equation}
	
	where $g(x)_i$ is the $i$-th output of the curve function at the point $x$ and $x_{\min}$ represents the vector of the target parameters.	
	
	%% initial parameters
	We set $\varepsilon_{\rel} := 3 \cdot 10^{-4}$ and $h := 10^{-14}$. These parameters required modifications in certain specific cases.
	We take the initial vector $x_0$ to be a vector of random numbers limited to the intervals $[0.9c_i,1.1c_i]$ where the $c_i$-s are the respective curve parameters.
	
	%% most important results, provide a diagram or table of improvement. Shouldn't take up too much space.
	We have experimented with many termination conditions and line search algorithms. Table \ref{tab:ore} summarizes the results when using the Backtracking Wolfe condition. Other conditions produce similar results (except for the number of function calls).
	
	\begin{table}[H]
	\centering
    \begin{tabular}{|l|c|c|c|c|c|c|}
    \hline
    \textbf{ORE Results} & \textbf{Time(ms)} & \textbf{Iterations} & \textbf{Value} & \textbf{Forward} & \textbf{Reverse} & \textbf{LS It.} \\ \hline
    \textbf{Before }         & 15213    & 65         & $7.18\cdot 10^{-19}$ & 81      & 1134    & 80 \\ \hline
    \textbf{AVX2}            & 9213     & 42         & $9.08\cdot 10^{-19}$ & 130     & 602     & 45 \\ \hline
    \textbf{AVX2  Polyfit}   & 4624     & 18         & $6.23\cdot 10^{-19}$ & 79      & 266     & 21 \\ \hline
    \textbf{AVX512}          & 17142    & 36         & $1.23\cdot 10^{-18}$ & 111     & 518     & 38 \\ \hline
    \textbf{AVX512  Polyfit} & 9627     & 17         & $5.69\cdot 10^{-19}$ & 71      & 252     & 18 \\ \hline
    \end{tabular}
    
    \caption{ORE Results} \label{tab:ore}
    \end{table}
    
    We can see a significant reduction in the number of iterations, reverse calls, and line search iterations.
    The final resulting value for the objective function only improves when we use polynomial regression.
    AVX-512 has a significant overhead cost thus not resulting in any speedup relative to AVX2.
    We can also observe that Polyfit allowed us to eliminate the need for extra forward function calls relative to the original implementation.
    
\subsection{Calibrating the LMM model} \label{lmm}

    %% Briefly explain the setup (just cite the paper for what AADC does here).
    %The LIBOR market model is a financial model of interest rates \cite{lmm_art}. 
	%We only need to know that it returns various outputs that we can correlate with real world data.

    We analyzed the simple example of calibrating the LMM model by a set of European options. In this particular case, the loss function can be expressed in the form 
    \begin{equation} \label{eq:lossexpect}
    G = \frac{1}{2} \sum_{i=1}^{m} (E y_i - C_i)^2.
    \end{equation}
    Applying AAD to a function of this form is not straightforward and was done in \cite{gl}.
	
	We employed a calibration tool provided by Matlogica (available at \cite{finalgit}) with the algorithm of \cite{gl} implemented that used Monte Carlo simulations, multithreading, LBFGS\texttt{++}, AVX, and AADC.
	This tool has two major functions: \verb*|simulate1()| and \verb*|simulate2()|.
	\verb*|Simulate1()| updates the values needed to calculate the function value at the chosen point.
	\verb*|Simulate2()| calls \verb*|simulate1()| to update the values and then computes the gradient at that point.
	Therefore, we can use \verb*|simulate1()| when we only want the function value (forward mode) and \verb*|simulate2()| to get the gradient (reverse mode).

    It was not possible to implement our SIMD modifications of Sections \ref{parlinesearch} and \ref{polyfit} due to the calibration tool already using AVX and multithreading. Implementing the changes would require a full refactoring of the tool. However, we split the interface into $3$ functions as explained in Section \ref{derivatives}. We also saved the reverse calls by deactivating the call of \verb*|simulate1()| when we called \verb*|simulate2()| in \verb*|getGrad()|. This could be done since the required outputs to compute the gradient were already calculated in \verb*|operator()|.
    
    %% Give initial parameters
    We set $\varepsilon_{\rel} := 10^{-6} $, $h:=10^{-5}$ and take the initial vector $x_0$ to be $90\%$ of the target parameters $ x_0 := 0.9x_{\min} $.
	
	We tested for unlimited LBFGS iterations and for LBFGS limited to 100 iterations (to reduce computational time). 
    We have also experimented with many termination conditions and line search algorithms. Table \ref{tab:lmm} summarizes the results when using the Bracketing Wolfe condition with only one thread and an unlimited number of LBFGS iterations.

    %% most important results, provide a diagram or table of improvement. Shouldn't take up too much space.
 
	\begin{table}[H]
	
	\centering
    \begin{tabular}{|l|c|c|c|c|c|c|}
    \hline
    \textbf{LMM Results}   & \textbf{Time(s)} & \textbf{Iterations} & \textbf{Value}   & \textbf{Forward} & \textbf{Reverse} & \textbf{LS It.} \\ \hline
    \textbf{AVX2 Before }  & 384     & 161        & $1.52\cdot 10^{-7}$ & $7.19\cdot 10^6$ & $3.59\cdot 10^6$ & 218 \\ \hline
    \textbf{AVX2 After}    & 338     & 133        & $1.53\cdot 10^{-7}$ & $7.65\cdot 10^6$ & $2.20\cdot 10^6$ & 189 \\ \hline
    \textbf{AVX512 Before} & 341     & 167        & $1.48\cdot 10^{-7}$ & $3.58\cdot 10^6$ & $1.79\cdot 10^6$ & 217 \\ \hline
    \textbf{AVX512 After}  & 261     & 117        & $1.49\cdot 10^{-7}$ & $3.26\cdot 10^6$ & $9.67\cdot 10^6$ & 160 \\ \hline
    \end{tabular}
    \caption{LMM Results} \label{tab:lmm}
    \end{table}

\section{Further improving the efficiency of LBFGS\texttt{++}} \label{concl}

% \textcolor{red}{5.1 with reference to parts of Algorithm \ref{bfgs} and of the structure of LBFGS\texttt{++}. Explain which changes to the Hessian update step don't work and which can}.

% \textcolor{green}{Hessian update: not really tested but I will try to remember and check}

    % A practitioner's idea was to allow the user to reuse the previous solver calculations to solve the same problem with a different initial point, so the initial Hessian approximation is given by the previous final matrix.
	% Although harder to implement, enabling the user to change the initial approximation of the Hessian to a full Hessian matrix calculated with another method, can be useful in some cases of applications.
	% Some simpler ideas were also tested but found to provide only improvements in some very rare circumstances.

    It would be interesting to try to improve the method of approximating the Hessian matrix in the case of the LBFGS version of Algorithm \ref{bfgs}. The LBFGS\texttt{++} implementation consists of a loop and it should be possible to improve efficiency by using SIMD to calculate portions of the loop in parallel. 
	
	We have considered the Nocedal-Wright line search algorithm \cite{nw} but have not been able to notably improve efficiency. Although we have implemented our modifications to this method it would be interesting to try to improve efficiency with specific changes to the interface.
	
	LBFGS-B is an extension of LBFGS to simple box constraints \cite{nw}. It would be interesting to see if our modifications are effective in this case. Extra verification may be needed to guarantee that we do not exit the box constraints.

    This research was undertaken before more effective methods for computing gradients of functions of the form \eqref{eq:lossexpect} were found in \cite{bgg} (expanding on the work of \cite{gl}). It would be interesting to implement the modifications of Section \ref{derivatives} for calibrating with these improved algorithms.

% \bibitem{andr} L. B.~Andersen, \textit{A Simple Approach to the Pricing of Bermudan Swaptions in the Multi-Factor Libor Market Model} (1999), \textcolor{violet}{\href{https://papers.ssrn.com/sol3/papers.cfm?abstract_id=155208}{SSRN 155208.}} 

% \bibitem{pit} V.~Piterbarg, \textit{A Practitioner's Guide to Pricing and Hedging Callable Libor Exotics in Forward Libor Models} (2003), \textcolor{violet}{\href{https://papers.ssrn.com/sol3/papers.cfm?abstract_id=427084}{SSRN 427084.}} 

% \bibitem{savin} B. N.~Huge, A.~Savine, \textit{LSM Reloaded - Differentiate xVA on your iPad Mini} (2017), \textcolor{violet}{\href{https://papers.ssrn.com/sol3/papers.cfm?abstract_id=2966155}{SSRN 2966155.}}  

% \bibitem{capr} L.~Capriotti, Y.~Jiang, A.~Macrina, \textit{AAD and least-square Monte Carlo: Fast Bermudan-style options and XVA Greeks},  Algorithmic Finance, 6(1-2) (2017), pp 35–49, \textcolor{violet}{\href{https://content.iospress.com/articles/algorithmic-finance/af201}{IOS Press.}}  

% \bibitem{giles} M.~Giles, \textit{An extended collection of matrix derivative results for forward and reverse mode automatic differentiation} (2008), \textcolor{violet}{\href{https://people.maths.ox.ac.uk/gilesm/files/NA-08-01.pdf}{M.Giles website report.}} 

% \bibitem{ggl}  A.~Goloubentsev, D.~Goloubentsev, E.~Lakshtanov, \textit{Adjoint Differentiation for generic matrix functions}, ArXiv e-prints (2021),  \textcolor{violet}{\href{https://arxiv.org/abs/2109.04913}{arXiv:2109.04913 [q-fin.CP]}}.

\end{document}